\documentclass[useAMS,usenatbib]{mn2e}
\usepackage{graphicx}
\title[Secondary resonances of co-orbital motions]
      {Secondary resonances of co-orbital motions}

\author[B. \'Erdi et al.]
       {B. \'Erdi,$^{1}$\thanks{E-mail: B.Erdi@astro.elte.hu}
        I. Nagy,$^{2}$
	Zs. S\'andor,$^{1}$
	\'A. S\"uli$^{1}$
	 and G. Fr\"ohlich$^{1}$   \\
	$^{1}$ Department of Astronomy, E\"otv\"os University, P\'azm\'any P\'eter s\'et\'any 1/A, 1117 Budapest, Hungary
	\\ 
	$^{2}$ Phyical Geodesy and Geodesical  
	       Research Group of the HAS, Technical 
	       University, M\H uegyetem rkp. 3, 1111 Budapest, Hungary }
	
\begin{document}

\date{Accepted 2007 July 7. Received 2007 June 21}

\pagerange{\pageref{firstpage}--\pageref{lastpage}} \pubyear{2007}

\maketitle

\label{firstpage}

\begin{abstract}
The size distribution of the stability region around the Lagrangian point $L_4$ is investigated in the elliptic restricted three-body problem as the function of the mass parameter and the orbital eccentricity of the primaries. It is shown that there are minimum zones in the size distribution of the stability regions, and these zones are connected with secondary resonances between the frequencies of librational motions around $L_4$. The results can be applied to hypothetical Trojan planets for predicting values of the mass parameter and the eccentricity for which such objects can be expected or their existence is less probable.
\end{abstract}


\section{Introduction}
The existence of Trojan planets in extrasolar planetary systems (EPS) is an interesting possibility. Some related problems have been recently studied in several papers. In the Solar system, more than 2000 Trojan asteroids have already been discovered. They move close to the orbit of Jupiter, near the stable Lagrangian points $L_4$ and $L_5$, and are in 1:1 mean motion resonance with the giant planet. Mars and Neptun also have co-orbital asteroidal companions, and there are co-orbital moons around Saturn as well. It can be expected that Trojan-type objects also exist in EPSs.
\par
So far there have been discussions on Trojan exoplanets. According to \citet{LaughlinandChambers2002}, two planets around a solar-mass star, with masses comparable to the mass of Jupiter or Saturn, can perform stable tadpole-type librations around the Lagrangian points. Pairs of Saturn-mass planets can also execute horseshoe orbits around a solar-mass star, but this is not possible for Jupiter-mass pairs. \citet{Nauenberg2002} determined numerically the non-linear stability domain of the triangular Lagrangian solutions in the general three-body problem as a function of the eccentricity of the orbits and the Routh's mass parameter. This study indicates that there is a wide range of Jupiter-size planetary masses and eccentricities for which such solutions could exist in EPSs.
\par
As to the formation of Trojan planets, several mechanisms have been proposed, including accretion from protoplanetary disk \citep{LaughlinandChambers2002, ChiangandLithwick2005} and capture into 1:1 resonance due to rapid mass growth of the planet \citep{ChiangandLithwick2005}, or convergent migration of multiple protoplanets \citep{Thommes2005,CresswellandNelson2006}. \citet{Beaugeetal2007} studied the formation of terrestrial-like Trojan planets from a swarm of planetesimals around the Lagrangian point $L_4$ of a giant planet. For the formation of the Trojan asteroids of Jupiter and Neptune, see \citet{Morbidellietal2005}. 
\par
Trojan exoplanets could be detected in several ways, like radial velocity (RV), astrometric, or transit photometry measurements. \citet{LaughlinandChambers2002} pointed out that a pair of comparable mass planets, both in tadpole and horseshoe orbits, would induce a characteristic pattern in the RV component of the central star that could be detected. Reanalysing the RV measurements of the systems HD 128311 and HD 82943, where two Jupiter-like planets are beleived to be in a 2:1 resonance, \citet{GozdziewskiandKonacki2006} demonstrated that it is possible to explain the variations of the RV of the hosting stars by a pair of Trojan planets in 1:1 resonance. According to \citet{GozdziewskiandKonacki2006}, this is also possible for the system HD 73526. However, in this case \citet{Tinneyetal2006} and \citet{Sandoretal2007} found only a 2:1 resonance.  \citet{FordandGaudi2006} developed a method for detecting Trojan companions to extrasolar planets by combining RV and photometric observations of transiting exoplanets. This method offers the potential to detect even terrestrial-mass Trojan planets using ground-based observations.
\par
Most of the known exoplanets are gaseous giants orbiting quite close to their hosting stars. The search for terrestrial-like planets is the ambitious aim of several ongoing and future ground-based and space projects (COROT has been in orbit since 2006 December, Kepler will be launched in 2008, others are planned for the next decade). It is a question of considerable interest, whether Earth-like planets exist in the habitable zone (HZ) of some exoplanetary systems. If there is a giant planet in the HZ of a system, the existence of another planet there is unprobable. However, as \citet{MenouandTabachnik2003} noted, terrestrial planets could exist near the stable Lagrangian points $L_4$ or $L_5$ of a giant planet moving in the HZ.
\par
\citet{ErdiandSandor2005} studied this possibility in detail, investigating five exoplanetary systems (HD 17051, HD 28185, HD 108874, HD 177830, and HD 27442) in which the only known giant planet moves in the HZ. By using the model of the elliptic restricted three-body problem (ERTBP), they numerically determined the region around $L_4$  of each system where stable tadpole-type librations are possible. Four other systems (HD 150706, HD 114783, HD 20367, and HD 23079) were also studied in which the orbit of the giant planet is partly outside the HZ due to its large eccentricity. It has been shown that in all studied systems there is an extended stability region around $L_4$, whose size depends on the mass ratio of the planet and the star and the orbital eccentricity of the giant planet. It is possible that Trojan exoplanets of negligible mass exist in these systems. Other dynamical studies \citep{Dvoraketal2004,Schwarzetal2005} also confirmed the possibility that small Trojan planets can exist in stable orbits with moderate eccentricities in these systems.
\par
\citet{Schwarzetal2007} reconsidered the system HD 108874, where a second giant planet was discovered \citep{Vogtetal2005} outside the HZ, perturbing strongly the inner planet. They showed by numerical integrations for $10^7$ yr that for most combinations of the orbital parameters, motions around $L_4$ are unstable. However, in some cases, terrestrial-like Trojan planets can exist in a small domain of stability. 
\par
Hypothetical Trojan planets in other multiple EPSs were also considered. 
\citet{Jietal2005} showed that Trojan planets with 0.1 -- 10 Earth-mass and with low eccentricities and inclinations could stay for $10^7$ yr at the triangular equilibrium points of the two massive planets of 47 UMa. According to \citet{Jietal2007}, Trojan planets with 0.01 -- 1 Earth-mass can also exist at least for $10^6$ yr in the triple planetary system HD 69830 in 1:1 mean motion resonance with the outer massive planet d.
\par
In this paper, we investigate how the size of the stability region around $L_4$ depends on the parameters of the problem. Since in more than 90\% of the EPSs discovered so far only one (giant) planet is known, we use a model of the three-body problem corresponding to a gravitational system of a star, a giant planet and a Trojan planet. We study cases where the Trojan planet has negligible mass and the orbit of the giant planet is elliptic (ERTBP), and also the Trojan planet can have finite mass and then we assume that the orbit of the giant planet is a perturbed ellipse. We determine the size of the stability region around $L_4$ depending on the mass of the giant planet (with respect to the mass of the star) or the masses of both planets and on the (initial) orbital eccentricity of the giant planet. For the sake of simplicity, here we only study coplanar orbits of the planets. 

\section{Maps of dynamical stability}
To determine the size of the stability region around $L_4$, we have to establish the dynamical character of orbits in the neighbourhood of this point.
In the model of the ERTBP, for this we used the Lyapunov characteristic indicators (LCI) \citep{Benettinetal1980, Froeschle1984}, computed for $10^4$ periods of the giant planet. During the numerical integration, we also checked the maximum eccentricity of the Trojan planet as another tool of dynamical characterization of orbits \citep{Dvoraketal2003}. 
We used dimensionless quantities with properly selected units. 
The sum of the masses of the star ($m_0$) and the giant planet ($m'$) was taken as unit mass. Then the mass of the star and the giant planet is $1-\mu$ and  $\mu$, respectively, where $\mu=m'/(m_0+m')$ is the mass parameter. The semi-major axis $a'$ of the orbit of the giant planet was taken as unit distance.
\par
We computed maps of dynamical stability depending on two parameters, $\mu$ and the eccentricity $e'$ of the orbit of the giant planet, that were changed in specific intervals. We changed $\mu$ between $5 \times 10^{-4}$ and $1.5 \times 10^{-2}$ in steps of $\Delta \mu=5 \times 10^{-4}$. (With a one solar-mass star this corresponds to a giant planet with about 0.5 -- 15 Jupiter-mass.) The eccentricity $e'$ was changed between 0 and 0.6 in steps of $\Delta e'=0.025$. For each possible pair of $\mu$ and $e'$, we computed a stability map showing the stable and unstable regions aroud $L_4$ in the parameter plane ($a,\lambda-\lambda'$), where $a$ is the semi-major axis of the orbit of the
Trojan planet and $\lambda-\lambda'$ is the synodic longitude, $\lambda$ and $\lambda'$ being the mean orbital longitude of the Trojan and the giant planet, respectively.
\begin{figure}
\includegraphics[width=0.90\linewidth]{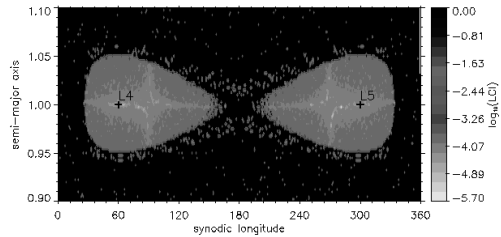}
\caption{The stability regions around $L_4$ and $L_5$ for $e'=0$ and $\mu=0.001$. There can be temporary stable horseshoe orbits outside the main stability regions where orbits have librational character.}
\end{figure}
\par
To obtain a stability map, we changed $a$ between 0.9 and 1.1 in steps of $\Delta a=0.0025$, and $\lambda$  between $0^{\circ}$ and $180^{\circ}$ in steps of $1^{\circ}$, taking the direction of the pericenter of the giant planet as starting point. The other initial conditions were: $\lambda'=0^{\circ}$, 
$e=0$, where $e$ is the eccentricity of the orbit of the Trojan planet. At each point of the ($a, \lambda-\lambda'$) plane, with the given initial data we computed the value of the LCI for the resulting orbit and represented its logarithm on a colour-coded scale. In what follows we discuss the main characteristics of the computed maps. Some representatives of them are shown in Figs 1--3. Light regions correspond to stable orbits (at least for the investigated time interval). The dark background corresponds to escape or collision orbits with the giant planet. Computations of the maximum eccentricity show that in such cases the eccentricity of the Trojan planet grows above 0.3.   \par
 Fig. 1 shows the stability regions around $L_4$  and $L_5$ for $e'=0$ (circular restricted three-body problem) and $\mu=0.001$. To compute this figure, we 
additionally changed $\lambda$  between $180^{\circ}$ and $360^{\circ}$. In the two main stability regions orbits have librational character. Outside the main stability regions there can be temporary stable horseshoe orbits. 
\par
Increasing the mass of the giant planet, the stability region becomes shorter in the synodic longitude and wider in the semi-major axis.
In the bottom panel of Fig. 2, obtained for $e'=0$, $\mu=0.002$, a ring structure appears near the edge of the stability region, corresponding to a high order resonance between the short and long period components of librational motions around $L_4$. Outside the main stability region, a chain of small islands can be seen, originating from the disruption of a former ring, due to the increase of the mass and increasing perturbations of the giant planet.
\begin{figure}
\includegraphics[width=0.90\linewidth]{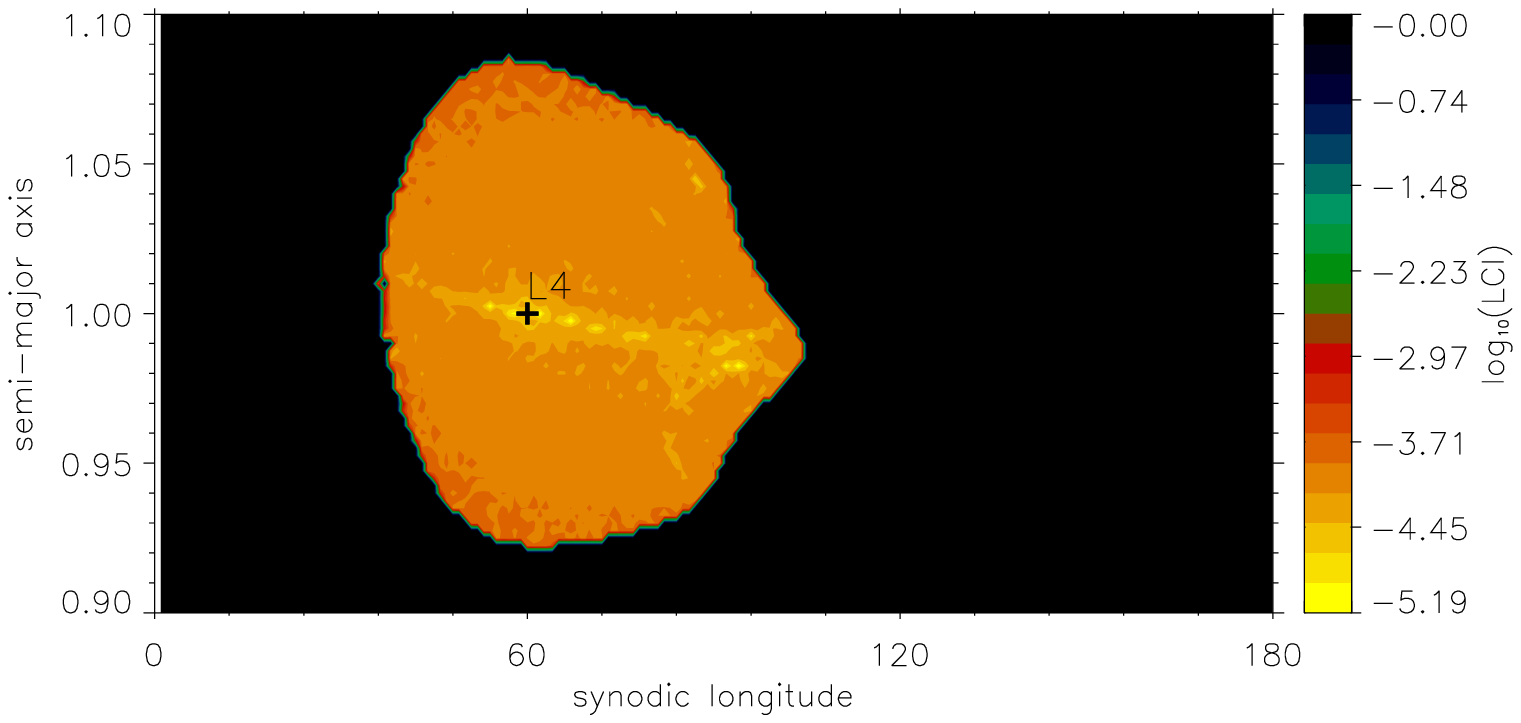}
\includegraphics[width=0.90\linewidth]{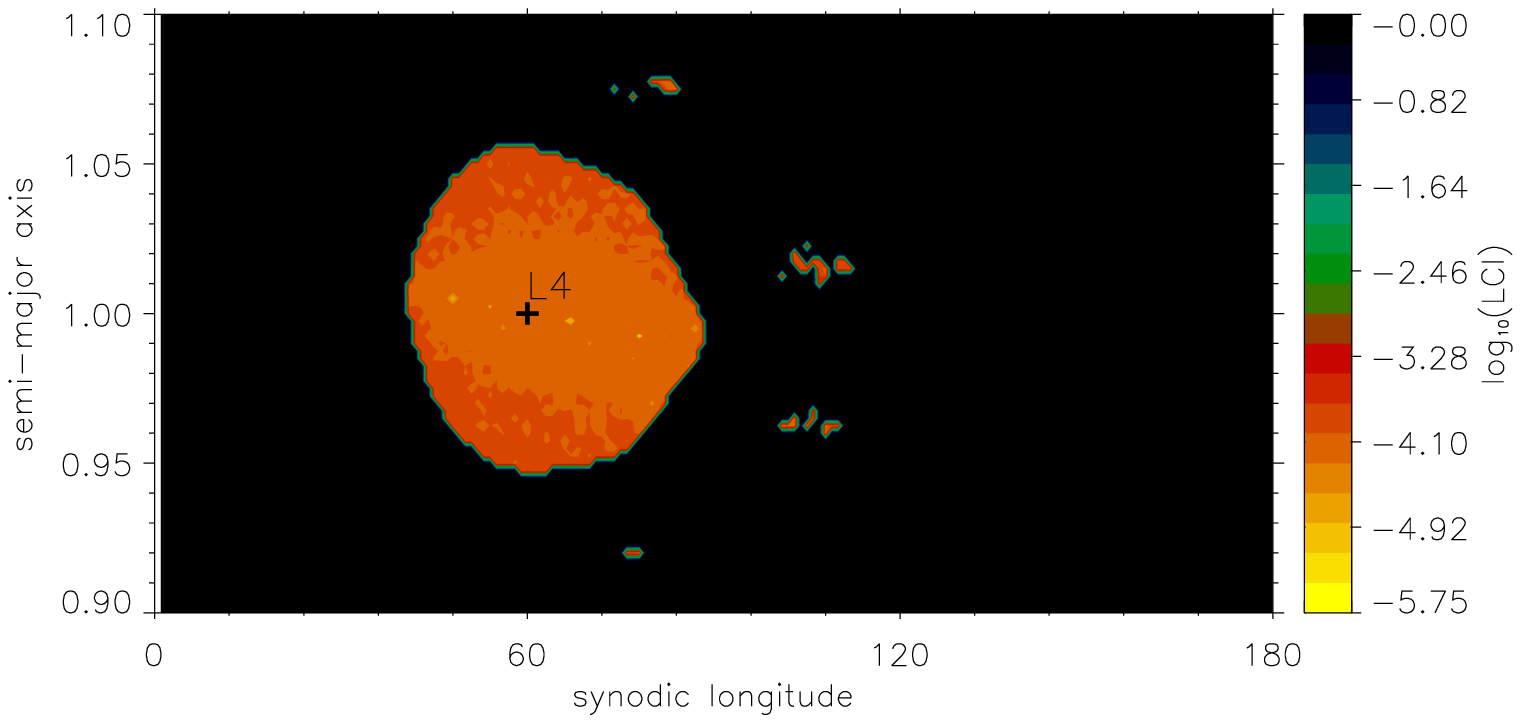}
\includegraphics[width=0.90\linewidth]{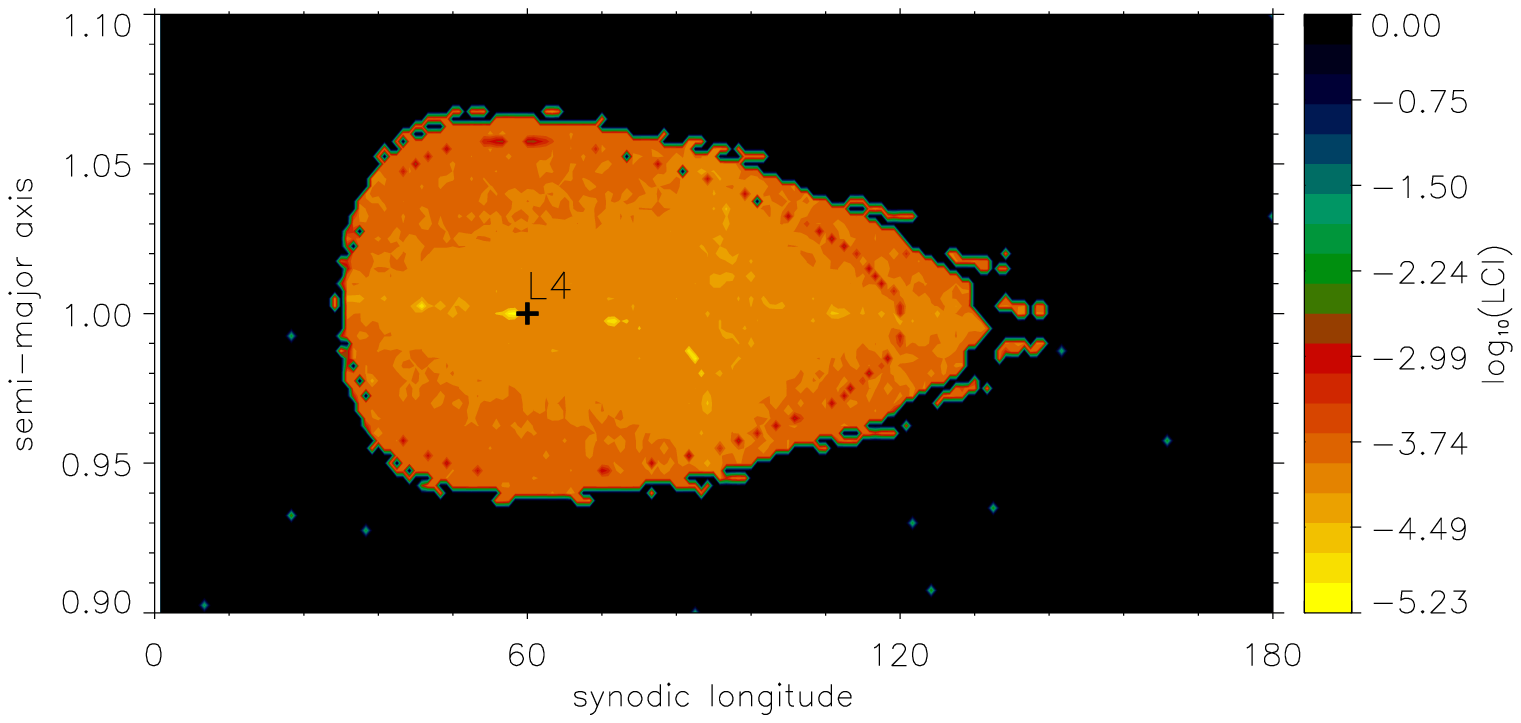}
\caption{The stability region around $L_4$ for $e'=0$ and $\mu=0.002$
(bottom panel), $\mu=0.006$ (middle panel ), and $\mu=0.007$ (top panel). Note the ring structure near the edge of the stability region in the bottom panel and the chain of small islands in the bottom and middle panels.}
\end{figure}
\par
One expects that with the increase of the mass of the giant planet, the extension of the stability region shrink due to larger perturbations. However, the comparison of the middle and top panels of Fig. 2 shows the opposite. In the former case ($e'=0$, $\mu=0.006$), the stability region is smaller than in the latter for larger $\mu$ ($e'=0$, $\mu=0.007$). This is one result of our computations: with the increase of the mass of the giant planet, the size of the stability region around $L_4$ does not decrease uniformly, smaller and larger regions alternate along the overall decreasing tendency. 
\par
This phenomenon can be also seen in the ERTBP. Fig. 3 shows for $e'=0.1$ the stability region for increasing values of $\mu$. Its size is decreasing as $\mu$ takes increasing values (bottom and middle panels), and it is larger again for a higher value of $\mu$ .  
\begin{figure}
\includegraphics[width=0.90\linewidth]{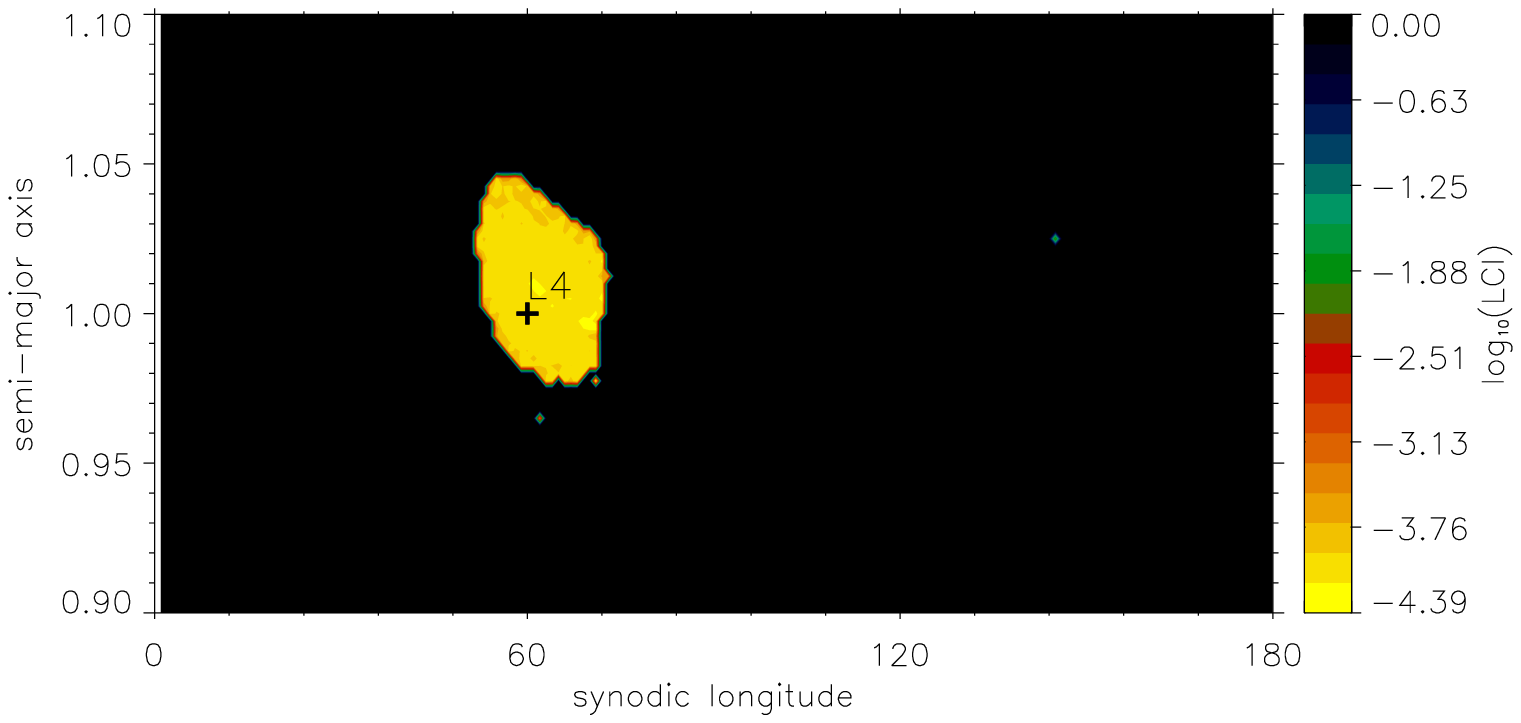}
\includegraphics[width=0.90\linewidth]{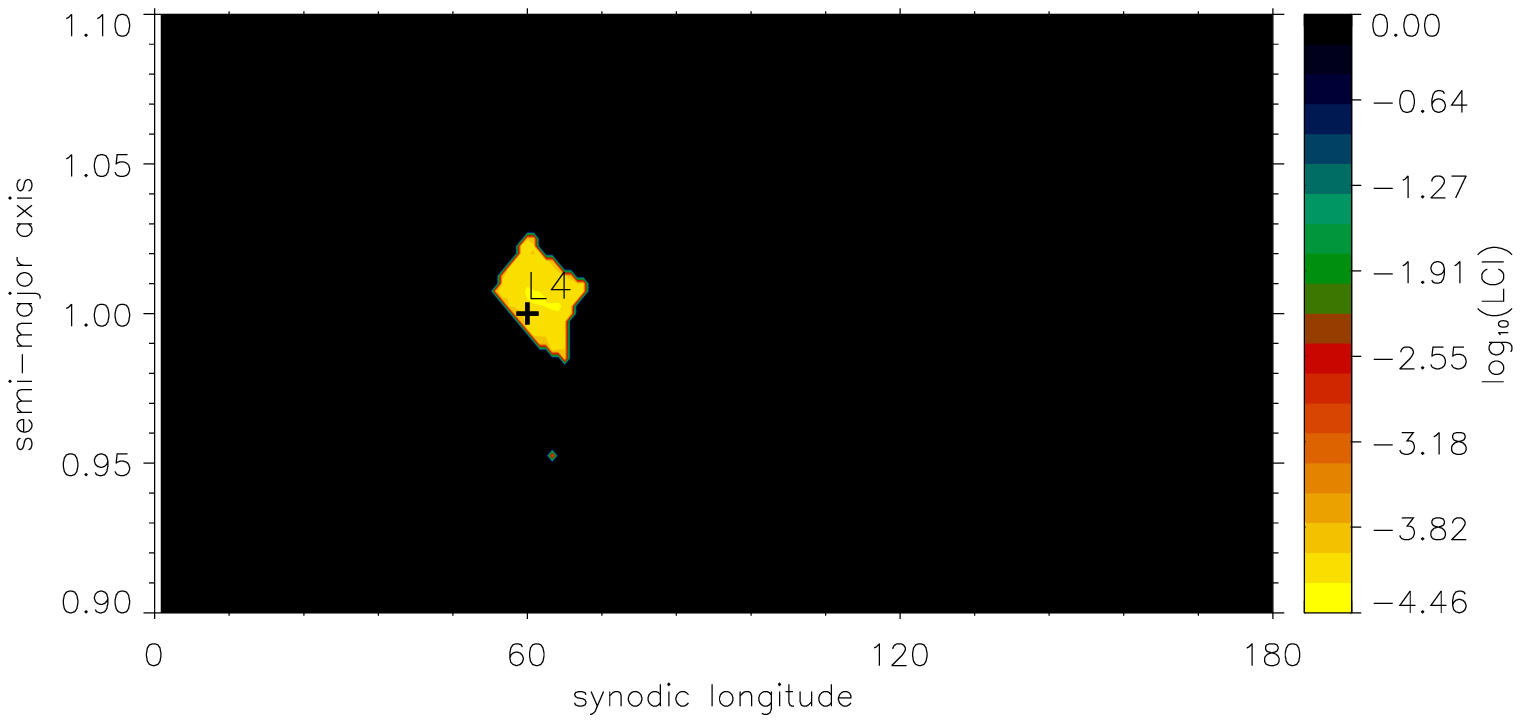}
\includegraphics[width=0.90\linewidth]{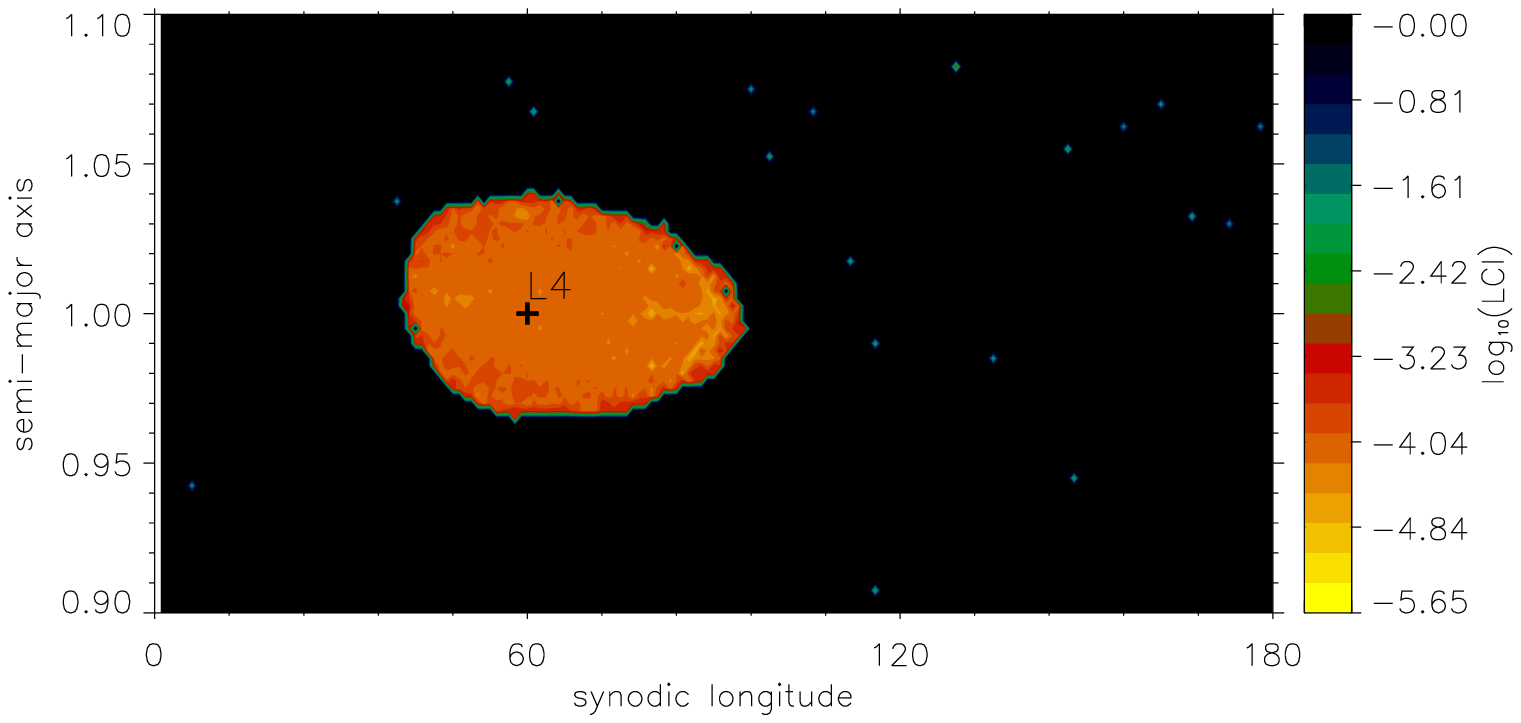}
\caption{The stability region around $L_4$ for $e'=0.1$ and $\mu=0.002$ (bottom panel), $\mu=0.007$ (middle panel), and $\mu=0.009$ (top panel).}
\end{figure}
\par
Studying the computed stability maps (not shown here), we can arrive at a similar conclusion: when we fix $\mu$ and change $e'$, with the increase of the eccentricity the decrease of the size of the stability region is not uniform.  
\par 
We also studied cases when the Trojan planet had non-negligible mass. Because of the larger complexity of the problem, in such cases we used the efficient method of the relative Lyapunov indicators (RLI) \citep{Sandoretal2000, Sandoretal2004} to determine the dynamical character of orbits around $L_4$. We computed stability maps on the ($a$, $\lambda-\lambda'$) plane by considering several values for the mass of the Trojan planet, corresponding to $m=1$, 10, and 100 Earth-mass ($m_{\mathrm{E}}$) and 1\ Jupiter-mass ($m_{\mathrm{J}}$), and for the mass of the giant planet, corresponding to 
$1, 2, \ldots, 7 \, m_{\mathrm{J}}$, assuming that the planets move around a solar-mass star. For the initial eccentricity of the giant planet, values were taken from the interval $e'=0-0.30$ in steps of 0.05. The other initial conditions were as before. 
\par
 The results of our computations show that the extension of the stability region do not change much under the effect of the mass of the Trojan planet. Even for $m=100\,m_{\mathrm{E}}$, the stability regions look like in Fig. 3 (for the corresponding values of $m'$ and initial $e'$). 
\par
We determined the stability regions around $L_4$ for the posssible combinations of the values of  $m$, $m'$ and $e'$ taken from the previously mentioned sets. The results show that for a given pair of $m'$ and $e'$, the size of the stability region does not change much with the increase of $m$. The changes are larger when $m$ is fixed, and either $m'$ or $e'$ is changed while the other is kept constant. The non-uniform decrease of the size of the stability region is observed in such cases as well. The computations confirm the existence of a stability region around $L_4$ even for $m=m_{\mathrm{J}}$, when the giant planet is of several Jupiter-masses and its orbit is very eccentric. As an example,  Fig. 4 shows the stability region for $m=1 \, m_{\mathrm{J}}$, $e'=0.3$, and $m'=1 \, m_{\mathrm{J}}$. Increasing $m'$ at this value of $m$ and $e'$, the size of the stability region decreases reaching a minimum at $m'=5\,m_{\mathrm{J}}$, after which it is larger again for $m'=6\,m_{\mathrm{J}}$, as the computations show. (Note, however, that in this case the stepsize is only $\Delta m'=1 m_{\mathrm{J}}$.)
\begin{figure}
\includegraphics[width=0.90\linewidth]{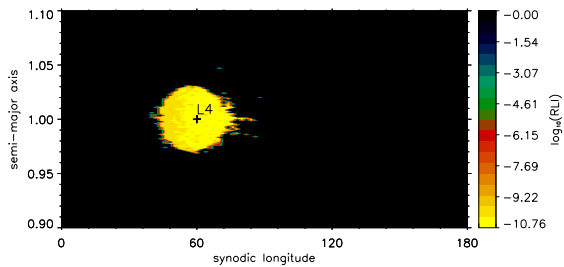}
\caption{The stability region around $L_4$ in the planar three-body problem for $m=1 \,m_{\mathrm{J}}$, $e'=0.3$, $m'=1\,m_{\mathrm{J}}$.}
\end{figure} 
\par
From the numerical investigations of this section, two main conclusions can be drawn: (i) the size of the stability region around $L_4$ depends essentially only on the mass parameter $\mu$ and the eccentricity $e'$; (ii) the variation of the size of the stability region is not uniform either with changing $\mu$ or $e'$.  To find an explanation for the second property needs a more thorough study, since though we considered many values of $\mu$ and $e'$, however, the applied resolution is not enough to get a satisfactory description of how the size of the stability region depends on these parameters. In the next section, therefore, we determine the distribution of the size of the stability region in a simplified model with high resolution as the function of $\mu$ and $e'$. This will enable us to study the fine structure of the size distribution of the stability region. 

\section{Size distribution of the stability region around $L_4$}
The extension of the stability region around the Lagrangian points $L_4$ and $L_5$ is an old problem of celestial mechanics that was studied in different contexts by many authors. Most of the papers refer to the problem of the stability of Jupiter's Trojan asteroids \citep*[see e.g.][]{Levisonetal1997, GiorgilliandSkokos1997, EfthymiopoulosandSandor2005, RobutelandGabern2006}.
\par
In the model of the ERTBP, the stability of the point $L_4$ depends on the mass parameter $\mu$ and the orbital eccentricity of the primaries which from now on we denote by $e$. The stability of orbits around $L_4$ also depends on these parameters. Non-linear stability of orbits around $L_4$ was studied by \citet{Gyorgyey1985} for the mass parameter of the Earth--Moon system and for a few values of the eccentricity. \citet{LohingerandDvorak1993} investigated the size of the stability region around $L_4$ by computing the number of stable orbits for a wide range of the values of $\mu$ and $e$. The most detailed survey on the size distribution of the stability region around $L_4$ depending on $\mu$ and $e$ was given by \citet{ErdiandSandor2005}.
\par
 Here we reconsider the problem by using a much higher resolution in $\mu$ and $e$ than in previous works. Fig. 5(a) shows the size distribution of the stability region around $L_4$ in the ERTBP depending on the mass parameter $\mu$ and eccentricity $e$. Note, that here and from now on $\mu$ has slightly different meaning than in Section 2. In order that we could compare our results with earlier works, we changed the system so that now the larger primary (the star) has a mass 1 and the smaller primary (the giant planet) has mass $\mu$ (thus their total mass is $1+\mu$, instead of 1). The test planet (the Trojan planet) has negligible mass.
\begin{figure*}
\centering
\includegraphics[width=\linewidth]{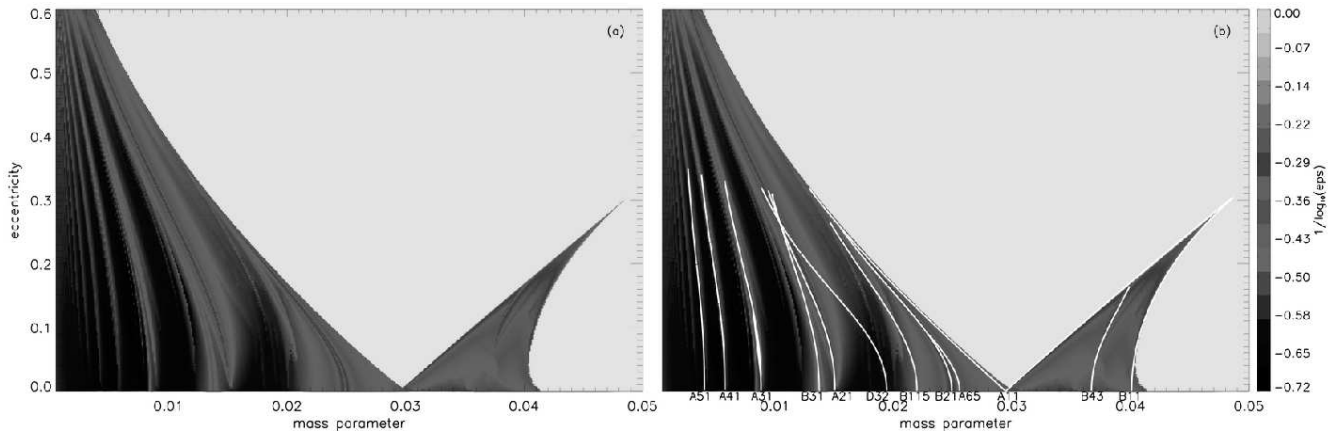}
\caption{(a) Size distribution of the stability region around $L_4$. (b) Secondary resonances of co-orbital motions.}
\end{figure*} 
\par
Fig. 5(a) was obtained as follows. For a given pair of the values of $\mu$ and $e$, we placed the test planet in the point $L_4$, and moving it outwards along the line connecting $L_4$ and the larger primary, we checked by numerical integration for $5 \times 10^4$ periods of the primaries the stability of the orbits, originating from consequtive points of this line with zero initial velocity with respect to the primaries. We defined stability in the simplest way: an orbit around $L_4$ was considered stable if it did not cross the line of the primaries. The reason for this definition was that when a crossing happens the orbit loses its librational character and we were interested in tadpole-type librations as possible orbits for Trojan planets. For any pair of $\mu$ and $e$, we determined the largest distance from $L_4$, denoted by $\varepsilon$, along the line of $L_4$ and the larger primary, until the orbit of the test planet starting from points of this line remains stable in the above sense. Certainly, the time interval in stability investigations is crucial. We know that the extension of the stability region around $L_4$ shrink with time \citep{Levisonetal1997}. We took $5 \times 10^4$ periods of the primaries for our stability investigations as a compromise. The general features of the size distribution of the stability region around $L_4$ as the function of $\mu$ and $e$ appear reliably during this time. On the other hand, the high resolution computations could be performed in a reasonable time. 
\par
 Changing $\mu$ and $e$ on a fine grid ($0.0001 \leq \mu \leq 0.05$, $\Delta \mu=0.0001$, $0 \leq e \leq 0.6$, $\Delta e=0.001$), we determined for each pair of $\mu$ and $e$ the limit value $\varepsilon$ (in the fraction of the initial distance of the primaries) corresponding to the largest stable librational orbit. The distance $\varepsilon$ can be thought as the approximate half thickness of the stability region described by the largest stable librational orbit. 
\par
Fig. 5(a) shows the distribution of $\varepsilon$ as the function of $\mu$ and $e$, that is the size distribution of the stability region around $L_4$ as the function of the parameters. For the sake of better visualization, instead of $\varepsilon$ the values of 1/log($\varepsilon$) are plotted in Fig. 5(a). The light region above the V-shaped boundary curve and on the right of the figure corresponds to instability, even the point $L_4$ is unstable for such values of $\mu$ and $e$. This boundary curve is well known from stability investigations of the point $L_4$ in the ERTBP \citep{Danby1964, Meire1981}. The point $L_4$ is stable and libration around $L_4$ is possible only below this curve. The novelty of Fig. 5(a) is, that in the region where $L_4$ can be stable, it indicates the size of the stable librational region around $L_4$. According to the colour code, darker domains correspond to more extended stability regions.
\par
As can be seen from Fig. 5(a), the size distribution of the stability region has a very complex structure. The size of the stability region is the largest when both $\mu$ and $e$ are small ($\mu<0.01$, $e<0.1$). The fine structure of the figure confirms our previous finding that the size of the stability region changes non-uniformly either fixing $e$ and varying $\mu$, or vice versa. It can be seen that there is also an extended stability region for small values of the eccentricity $(e<0.1)$ between $\mu=0.014-0.020$. 
\par
The dominant features of Fig. 5(a) are the dark and light stripes, forming up finger-like structures on the left of the figure. One might suspect that the light stripes, indicating less extended stability regions, are related to resonances between the frequencies of librational motions. In the next section, we consider this possibility and see how resonances of librational motions around $L_4$ in the ERTBP can be determined.
\par
We also computed the size distribution of the stability region around $L_4$ by giving masses to the Trojan planet and we found that the structure shown in Fig. 5(a) remains even for a Trojan planet as large as 1 Jupiter-mass. The only difference is that with the increase of the mass of the Trojan planet, the stripes of the figure are slightly shifted toward the direction of lower mass parameters. Also, there is only a slight difference in the size distributions corresponding to pericentric or apocentric (as in Fig. 5a) initial positions of the primaries. Thus the structure of Fig. 5(a) seems to be quite general.

\section{Resonances of librational motions}
In the circular restricted three-body problem, motions around $L_4$ has two frequencies, $n_{\rmn{s}}$ and $n_{\rmn{l}}$, corresponding to the short and long period components of libration. In the ERTBP, however, the point $L_4$ has an elliptic motion itself, similar to the elliptic motion of the primaries, and a corresponding mean motion $n$. This frequency combines with the frequencies of the short and long period librations, and thus motions around $L_4$ have four frequencies (and linear combinations of them) which for $e \rightarrow 0$ approach the limit values
$n_{\rmn{s}}$, $n-n_{\rmn{l}}$, $n_{\rmn{l}}$, and $n-n_{\rmn{s}}$.
Taking $n$ as unit frequency, the normalized frequencies $n_{\rmn{s}}$ and $n_{\rmn{l}}$ are given by \citep{Rabe1970,Rabe1973}:
\begin{equation}
 n_{\rmn{s}}^2=\frac{1}{2} (1+ \sqrt{1-4\kappa} ), \quad 
 n_{\rmn{l}}^2=\frac{1}{2} (1- \sqrt{1-4\kappa} ),
\end{equation}
where
\begin{equation}
   \kappa=\frac{27}{4}\frac{\mu}{(1+\mu)^4}.
\end{equation} 
The frequencies $n_{\rmn{s}}$ and $n_{\rmn{l}}$ are real until $\mu \leq 0.04006\ldots$, and when the equality is held, $n_{\rmn{s}}=n_{\rmn{l}}=1/\sqrt{2}$ is a double root. (Note that values of $\mu$ of this section can be simply transformed to the system, where the primaries have masses $1-\mu$ and $\mu$, dividing them by $1+\mu$. This results in the familiar $0.03852\ldots$ value for the critical mass parameter.) 
\par
The normalized frequencies $n_{\rmn{s}}$, $1-n_{\rmn{l}}$, $n_{\rmn{l}}$, and $1-n_{\rmn{s}}$ are given as the function of the mass parameter $\mu$ in Fig. 6. The frequency curves $n_{\rmn{s}}(\mu)$ and $n_{\rmn{l}}(\mu)$ meet at the point B, the curves $1-n_{\rmn{l}}$ and $1-n_{\rmn{s}}$ at the pont C, both for $\mu=0.04006$, and the common frequencies are $1/\sqrt 2$ and $1-1/\sqrt 2$, respectively. The curves $1-n_{\rmn{l}}$ and $n_{\rmn{l}}$ cross at the point A for $\mu=0.02944$ and the common frequency is 1/2. The points A, B, and C correspond to 1:1 resonances between the relevant frequencies.
\par
In the ERTBP, the four frequencies of librations around $L_4$ can be determined from an equation derived by \citet{Rabe1973}, which relates the frequency $z$, the modified mass parameter $\kappa$ (see Equation 2) and the eccentricity $e$: 
\begin{equation}
   D(z,\kappa,e)=A(z,\kappa)+e^2 B(z,\kappa)+e^4 C(z,\kappa)=0.
\end{equation}
The functions $A(z,\kappa)$, $B(z,\kappa)$, and $C(z,\kappa)$ are given in the Appendix. Equation (3) is 20th order in $z$, 5th order in $\kappa$, and 4th order in $e$. Still, this is an approximate equation, obtained from an infinite determinant of an infinite number of homogeneous linear equations for the coefficients of oscillating solutions around $L_4$. Nevertheless, Equation (3) can be used to determine the frequencies for any given $\kappa$ (corresponding to $0 < \mu \leq 0.05$) and for small and moderate values of $e$. \citet{Rabe1973} used Equation (3) for a fourth-order stability analysis of $L_4$ in the ERTBP, and to determine the transition curve between stability and instability in the $\kappa,e$ plane.   
\begin{figure}
\includegraphics[width=0.80\linewidth]{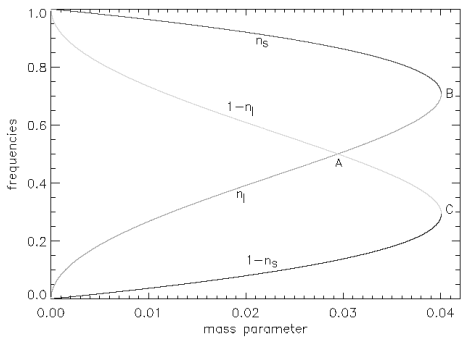}
\caption{Limit frequencies of librational motions in the elliptic restricted three-body problem for $e \rightarrow 0$.}
\end{figure}
\par
In general, Equation (3) has four different roots for $z$ between 0 and 1, $z=1$ corresponding to the unit frequency of the primaries. \citet{Rabe1973} determined two solutions for Equation (3): 
\begin{enumerate}
\item  $e=0$, $\kappa=3/16$ ($\mu=0.02944$), $z=1/2$; 
\item  $e=0$, $\kappa=1/4$ ($\mu=0.04006$), $z=1/\sqrt{2}$. 
\end{enumerate}
The solution (i) corresponds to the point A in Fig. 6, while the solution (ii) to the point B. These are critical points, where two frequencies are equal, and in the $\kappa,e$ plane boundary curves between stability and instability are defined by these two fixed frequencies. Starting from these solutions, \citet{Rabe1973} derived approximate analytical solutions from  Equation (3) for the boundary curves in the form of series expansions, valid for moderate values of $e$. 
\par 
For $z=1/2$, Rabe's approximate solution is of third-order: 
\begin{equation}
   \kappa=\frac{3}{16} \pm \frac{\sqrt{33}}{16}e+0.04165 e^2
                       \mp 0.14926 e^3.
\end{equation}
Adding to this a numerically determined fourth-order term, $0.07e^4$,  
the solution (4) is shown in Fig. 5(b) as the curve labelled by A11. As can be seen, it gives quite well the V-shaped boundary curve between the stable and unstable region of $L_4$ for moderate eccentricities ($e< 0.3$). Along this curve, $1-n_{\rmn{l}} =n_{\rmn{l}}=1/2$, thus the boundary curve is the place where there is a 1:1 resonance between the frequencies 
$1-n_{\rmn{l}}$ and $n_{\rmn{l}}$, and moreover a 2:1 resonance between the frequencies 1 (the frequency of the primaries) and $n_{\rmn{l}}$.
\par
For $z=1/\sqrt{2}$, \citet{Rabe1973} derived a fourth-order solution. This solution gives the right side boundary between the stable and unstable region of $L_4$ in the $\kappa,e$ plane. However, it seems that the fourth-order term is erroneous in Rabe's solution. For this boundary \citet{Tschauner1971} derived an exact analytical solution. Using Tschauner's solution, Rabe's solution can be corrected, resulting in:
\begin{equation}
   \kappa=\frac{1}{4}+\frac{1}{2}e^2+\frac{9}{32}e^4.
\end{equation}    	       
This solution is shown in Fig. 5(b) as the curve B11. Along this curve, 
$n_{\rmn{s}}=n_{\rmn{l}}=1/\sqrt{2}$, thus this boundary curve is the place where there is a 1:1 resonance between the frequencies $n_{\rmn{s}}$ and 
$n_{\rmn{l}}$. As can be seen, the theoretical B11 curve fits very well the numerically obtained boundary, except for low values of the eccentricity ($e<0.05$). We do not have an explanation for this yet.
\par
Apart from the above two solutions, no other solution was given to Equation (3). This might be due to the algebraic difficulties which \citet{Rabe1970} noted in connection with an earlier, simpler form of the equation. However, Equation (3) offers the possibility to compute the frequencies for given mass parameters and eccentricities. We suspect that the structures in Fig. 5(a) are connected with resonances between the frequencies. We have already seen that the boundary curves between the stable and unstable regions of $L_4$ are places of 1:1 resonances. Considering the frequency curves in Fig. 6, several other types of resonances are possible. Table 1 lists the possible pairs of the frequencies with a suggested name for the type of the resonance.
\begin{table}
\caption{Types of resonances}
\begin{tabular}{cc}
\hline
type & resonance \\
 \hline
 A &  $(1-n_{\rmn{l}}):n_{\rmn{l}}$     \\
 B &  $n_{\rmn{s}}:n_{\rmn{l}}$        \\
 C &  $(1-n_{\rmn{l}}):(1-n_{\rmn{s}})$ \\
 D &  $n_{\rmn{s}}:(1-n_{\rmn{l}})$     \\
 E &  $n_{\rmn{s}}:(1-n_{\rmn{s}})$     \\
 F &  $n_{\rmn{l}}:(1-n_{\rmn{s}})$     \\
 \hline
\end{tabular}   
\end{table}
\par
Equation (4) is an approximate solution for the A 1:1 resonance, while Equation (5) is for the B 1:1 resonance (that is why they are labelled as A11 and B11 in Fig. 5b). As can be seen from Fig. 6, there is also a C 1:1 resonance with frequency $1-1/\sqrt{2}$. It was shown by \citet{Rabe1970} that due to the structure of Equation (3), if $z$ is a solution, so is $1-z$. Since Equation (5) is a solution for $z=1/\sqrt{2}$, thus it is also a solution for $z=1-1/\sqrt{2}$. Therefore, Equation (5) is also an approximate  solution for the C 1:1 resonance. 
\par
Numerically, a 1:1 resonance solution can be determined from Equation (3) by fixing the frequency at the resonant value and solving the equation for $e$ with a given $\kappa$, and repeating the procedure for other values of $\kappa$. Thus an $e(\kappa)$ curve is obtained corresponding to the 1:1 resonance. Since Equation (3) is fourth-order in $e$, reducible to second order, the solution is easy.
\par
Solutions for other resonances can be obtained in the following way. If there are two frequencies in resonance, say $z$ and $rz$, where $r$ is the ratio of the resonance, then both frequencies have to satisfy Equation (3). Then we have
two equations
\begin{equation}
  D(z,\kappa,e)=0, \quad  D(rz,\kappa,e)=0.
\end{equation}
For a given value of $\kappa$, we express the solution for $e$ from both equations, containing $z$ as parameter, and search for such value of $z$, for which the two solutions for $e$ become equal. (In the formula for the solution of the second order equation giving $e$, the square root must be taken with  negative sign to obtain good solution.) Repeating the procedure for other values of $\kappa$, the curve $e(\kappa)$ of the resonance $r$ is obtained. Note that the resonant frequency is changing along this curve. 
\par
Using the above method, we computed several resonant solutions. The corresponding resonant curves are shown in Fig. 5(b). Table 2 gives the initial points of these curves (mass parameter $\mu$ at $e=0$) and one of the initial resonant frequencies (dividing it by the resonant ratio, the other frequency can be obtained). Table 2 also contains the initial points of the solutions A11, B11 given by Equations (4) and (5). Since Equation (3) is fourth-order in $e$, it gives satisfactory solution until about $e<0.3$. Thus we computed the resonant curves below this limit.
\begin{table}
\caption{Initial points of the resonant curves at $e=0$}
\begin{tabular}{ccccc}
\hline
type & resonance & frequency &   $\mu$ \\
\hline
 A   &  5:1      &  0.83333333      &   0.00403325  \\
 A   &  4:1      &  0.80000000      &   0.00575455  \\
 A   &  3:1      &  0.75000000      &   0.00883461  \\
 B   &  3:1      &  0.94868330      &   0.01370120  \\
 A   &  2:1      &  0.66666667      &   0.01507644  \\
 D   &  3:2      &  0.92307692      &   0.01940524 \\
 B   & 11:5 \,   &  0.91036648      &   0.02195743   \\
 B   &  2:1      &  0.89442719      &   0.02489879   \\
 A   &  6:5      &  0.54545455      &   0.02554130   \\
 A   &  1:1      &  0.50000000      &   0.02943725   \\
 B   &  4:3      &  0.80000000      &   0.03668353   \\
 B   &  1:1      &  0.70710678      &   0.04006421   \\
\hline 
\end{tabular}
\end{table}
\par
In Fig. 5(b), the two most prominent minimum zones are connected to two 2:1 resonances, A 2:1 and B 2:1. The B21 resonant curve goes sharply in the middle of the minimum zone, which is bordered by the curve of the A 6:5 resonance. 
The minimum zones on the left of the figure are connected mainly with high order A type resonances.
\par
Close to the A21 resonant curve, there goes the B31 curve and they meet at the point $\mu_{\rmn{c}}=0.0103576$, $e_{\rmn{c}}=0.265615$. A third resonant curve, D32 goes exactly through this point, indicating multiple resonances. Noting, that the frequencies change along a resonant curve, it is convenient to denote by $z_{\rmn{s}}$ the frequency whose limit frequency is $1-n_{\rmn{l}}$. Then the three frequencies at the crossing point of the three resonant curves are:
$n_{\rmn{s}}=0.9629994$, $n_{\rmn{l}}=0.3209998$, $z_{\rmn{s}}=0.6419995$. Indeed, $n_{\rmn{s}}/n_{\rmn{l}}=3/1$, $z_{\rmn{s}}/n_{\rmn{l}}=2/1$, and $n_{\rmn{s}}/z_{\rmn{s}}=3/2$. There is also a Laplace-type relation between the three frequencies:  $z_{\rmn{s}}+n_{\rmn{l}}-n_{\rmn{s}}=0$. Computations show that the size of the stability region around $L_4$ for the $\mu_{\rmn{c}}$ and $e_{\rmn{c}}$ value of the cross point is very reduced, it is less than $2^{\circ}$ in synodic longitude.
\par
The starting points of the B31 and B21 resonant curves are at 
$\mu=0.01370120$ and $\mu=0.02489879$, corresponding to
0.013516016 and 0.024293897 in the usual mass parameter system. These are the values for which  \citet{Deprit1967} showed the non-linear instability of $L_4$ in the circular restriced three-body problem. Interestingly, as can be seen from Fig. 5, along the A21 resonant curve the size of the stability region around $L_4$ seems less extended than along the nearby B 3:1 resonance, though $L_4$ is non-linearly stable at its $e=0$ starting point.  
\par
On the right of the figure, the most prominent minimum zone is connected with the B 4:3 resonance. Interestingly, this zone begins at $e \approx 0.1$ and not at $e=0$. 
\par
It is worth to compare Fig. 5(a) with the results described in Section 2 concerning the maps of dynamical stability. For each pair of the mass parameter and the giant planet's eccentricity, we computed the LCI for about $1.5\times 10^4$ orbits by changing the semi-major axis and the orbital longitude of the Trojan planet. From this we can  calculate the ratio of the number of stable orbits with respect to all the investigated orbits. This ratio can be assigned to each pair of $\mu$ and $e$, and in this way a map can be obtained showing the ratio of stable orbits. The results are shown in Fig. 7 (note that we extended the computations of Section 2 up to $\mu=0.05$). Though the resolution of this figure is much less than in Fig. 5(a) (and the mass parameters are slightly different), still the basic structure is similar. The minimum zones, identified in Fig. 5(b) as the A 2:1, A 3:1 and A 4:1 resonances, appear also in Fig. 7. The zone of the A 2:1 resonance at $\mu \approx 0.015$ indicates an especially sharp minimum. Fig. 7 also comfirms the more probable occurance of stable orbits around $L_4$ for small values of the mass parameter and the eccentricity. We note that Fig. 7 is in good agreement with Fig. 6 of \citet{ErdiandSandor2005} obtained by the method of the RLI. 
\begin{figure}
\centering
\includegraphics[width=0.9\linewidth]{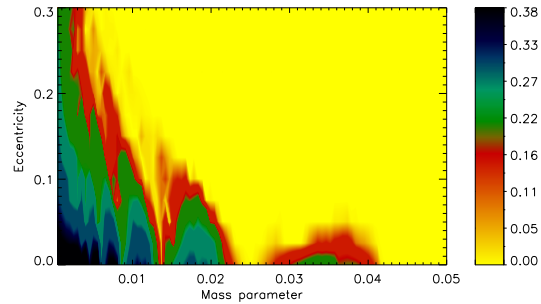}
\caption{Ratio of stable orbits depending on the mass parameter and eccentricity.}
\end{figure}

\section{Conclusions}
In studying the possibility of Trojan exoplanets, the size of the stability region around the Lagrangian points $L_4$  and $L_5$ is crucial. We determined the size distribution of the stability region around $L_4$ in the elliptic restricted three-body problem depending on the mass parameter and the eccentricity. The size distribution has a fine structure with several minimum zones. We have shown that these minimum zones are connected with resonances between the frequencies of librational motions around $L_4$. The situation is analogous to the Kirkwood gaps at mean motion resonances in the distribution of the main belt asteroids of the Solar system.
\par
One main minimum zone, the region of the A 2:1 resonance is situated around $\mu=0.015$, which corresponds to a celestial body of about 15 Jupiter-mass around a solar-mass star. This could be called brown dwarf limit in the sense that it is not probable that brown dwarfs of this size can possess Trojan companions due to the reduded size of the stability region. However, the stable region above the brown dwarf limit in Fig. 5 suggests the possibility of Trojan planets accompanied to larger objects. The regions of the higher order A 5:1, A 4:1, and A 3:1 resonances also indicate mass parameter values (0.004, 0.006, 0.009) and correspondig eccentricities for which single planet exoplanetary systems  hardly can have Trojan planets. Circumstances are most favorable for Trojan planets to exist, when both the mass parameter and the giant's planet eccentricity are small. However, higher order resonances may also affect this region. 

\label{lastpage}

\section*{Acknowledgments}
The support of the Hungarian Scientific Research Fund under the grants OTKA T043739 and D048424 is acknowledged. Parts of the numerical computations were done on the NIIDP supercomputer in Hungary.

\onecolumn

\appendix

\section{Coefficients of Rabe's frequency equation}

\begin{eqnarray}
& & A=576z^4-2800z^6+5596z^8-5925z^{10}+3581z^{12}-1250z^{14}+246z^{16}
   -25z^{18}+z^{20} \nonumber \\
  & & \quad \quad +\kappa\,(-576z^2+880z^4-28z^6-295z^8+24z^{10}+10z^{12}-20z^{14}+5z^{16})
       \nonumber \\ 
  & & \quad \quad +\kappa^2(1200z^2-1860z^4+230z^6+330z^8+90z^{10}+10z^{12}) 
      \nonumber \\
  & & \quad \quad +\kappa^3(144-20z^2+590z^4+140z^6+10z^8) \nonumber \\
  & & \quad \quad +\kappa^4(24+55z^2+5z^4)+\kappa^5,
\end{eqnarray}

\begin{eqnarray}
 & & B=13824z^4-58992z^6+102936z^8-95319z^{10}+51168z^{12}-16386z^{14}
        +3060z^{16}-303z^{18}+12z^{20} \nonumber \\
 & & \quad \quad
    +\kappa\,(-15840z^2+14704z^4+3058z^6+1156z^8-4456z^{10}+1856z^{12}
              -562z^{14}+84z^{16})  \nonumber \\
 & & \quad \quad 
  +\kappa^2(33208z^2-27678z^4+2124z^6+4238z^8+888z^{10}+180z^{12}) \nonumber \\
 & & \quad \quad 
 +\kappa^3(3912-1370z^2+7112z^4+1674z^6+156z^8) \nonumber \\
 & & \quad \quad
  +\kappa^4(326+527z^2+48z^4),
\end{eqnarray}

\begin{eqnarray}
 & & C=71136z^4-336122z^6+\frac{1249017}{2}z^8-\frac{1188315}{2}z^{10}
        +\frac{2551517}{8}z^{12}-\frac{202743}{2}z^{14}+\frac{75465}{4}z^{16}
	-1874z^{18}+\frac{597}{8}z^{20} \nonumber \\
 & & \quad \quad
   +\kappa \left(-153720z^2+118811z^4+\frac{166205}{4}z^6
                 +\frac{272839}{8}z^8-\frac{234923}{4}z^{10}
		 +22918z^{12}-\frac{11251}{2}z^{14}
		 +\frac{5529}{8}z^{16} \right) \nonumber \\
 & & \quad \quad
   +\kappa^2 \left(31392+285673z^2-208906z^4+\frac{133927}{8}z^6
                  +\frac{248663}{8}z^8+\frac{37293}{8}z^{10}
		  +\frac{11709}{8}z^{12} \right) \nonumber \\	
 & & \quad \quad
   +\kappa^3 \left(33952+5564z^2+\frac{361051}{8}z^4+\frac{39175}{4}z^6
                   +\frac{7275}{8}z^8 \right) \nonumber \\
 & & \quad \quad
    +\kappa^4 \left(\frac{1589}{4}+\frac{5645}{8}z^2+\frac{249}{4}z^4 \right).		
\end{eqnarray}


\begin{thebibliography}{99}

\bibitem[\protect\citeauthoryear{Beaug\'e et al.}{2007}]{Beaugeetal2007}
    Beaug\'e C., S\'andor Zs., \'Erdi B., S\"uli \'A., 2007, A\&A, 463, 359
    
\bibitem[\protect\citeauthoryear{Benettin et al.}{1980}]{Benettinetal1980}
   Benettin G., Galgani L., Giorgilli A., Streclyn J.M., 1980, Meccanica, 15, 2

\bibitem[\protect\citeauthoryear{Chiang \& Lithwick}{2005}]
     {ChiangandLithwick2005}
      Chiang E.I., Lithwick Y., 2005, ApJ, 628, 520 

\bibitem[\protect\citeauthoryear{Cresswell \& Nelson}{2006}]
        {CresswellandNelson2006}
	Cresswell P., Nelson R.P., 2006, A\&A, 450, 833  

\bibitem[\protect\citeauthoryear{Danby}{1964}]{Danby1964}
       Danby J.M.A, 1964, AJ, 69, 165

\bibitem[\protect\citeauthoryear{Deprit \& Deprit--Bartholome}{1967}]{Deprit1967}
       Deprit A., Deprit--Bartholome A., 1967, AJ, 72, 173

\bibitem[\protect\citeauthoryear{Dvorak et al.}{2003}]{Dvoraketal2003}
Dvorak R., Pilat--Lohinger E., Funk B., Freistetter F., 2003, A\&A, 410, L13

\bibitem[\protect\citeauthoryear{Dvorak et al.}{2004}]{Dvoraketal2004}
Dvorak R., Pilat--Lohinger E., Schwarz R., Freistetter F., 2004, A\&A, 426, L37

\bibitem[\protect\citeauthoryear{Efthymiopoulos \& S\'andor}{2005}]
          {EfthymiopoulosandSandor2005}
	  Efthymiopoulos C., S\'andor Zs., 2005, MNRAS, 364, 253 

\bibitem[\protect\citeauthoryear{\'Erdi \& S\'andor}{2005}]{ErdiandSandor2005}
       \'Erdi B., S\'andor Zs., 2005, CeMDA, 92, 113

\bibitem[\protect\citeauthoryear{Ford \& Gaudi}{2006}]{FordandGaudi2006}
     Ford E.B., Gaudi B.S., 2006, ApJ, 652, L137        

\bibitem[\protect\citeauthoryear{Froeschl\'e}{1984}]{Froeschle1984}
    Froeschl\'e Cl., 1984, CeM, 34, 95

\bibitem[\protect\citeauthoryear{Giorgilli \& Skokos}{1997}]
         {GiorgilliandSkokos1997}
	 Giorgilli A., Skokos C., 1997, A\&A, 317, 254

\bibitem[\protect\citeauthoryear{Go\'zdziewski \& Konacki}{2006}]
        {GozdziewskiandKonacki2006}
	Go\'zdziewski K., Konacki M., 2006, ApJ, 647, 573

\bibitem[\protect\citeauthoryear{Gy\"orgyey}{1985}]{Gyorgyey1985}
       Gy\"orgyey J., 1985, CeM, 36, 281

\bibitem[\protect\citeauthoryear{Ji et al.}{2005}]{Jietal2005}
   Ji J., Liu L., Kinoshita H., Li G., 2005, ApJ, 631, 1191

\bibitem[\protect\citeauthoryear{Ji et al.}{2007}]{Jietal2007}
  Ji J., Kinoshita H., Liu L., Li G., 2007, ApJ, 657, 1092

\bibitem[\protect\citeauthoryear{Laughlin \& Chambers}{2002}]
       {LaughlinandChambers2002}
       Laughlin G., Chambers J.E., 2002, AJ, 124, 592

\bibitem[\protect\citeauthoryear{Levison, Shoemaker \& Shoemaker}{Levison et al.}{1997}]{Levisonetal1997}
	Levison H., Shoemaker E., Shoemaker C., 1997, Nat, 385, 42 

\bibitem[\protect\citeauthoryear{Lohinger \& Dvorak}{1993}]
          {LohingerandDvorak1993}
     Lohinger E., Dvorak R., 1993, A\&A, 280, 683

\bibitem[\protect\citeauthoryear{Meire}{1981}]{Meire1981}
     Meire R., 1981, CeM, 23, 89

\bibitem[\protect\citeauthoryear{Menou \& Tabachnik}{2003}]
       {MenouandTabachnik2003}
       Menou K., Tabachnik S., 2003, ApJ, 583, 473
 
\bibitem[\protect\citeauthoryear{Morbidelli et al.}{2005}]{Morbidellietal2005}
 Morbidelli A., Levison H.F., Tsiganis K., Gomes R., 2005, Nat, 435, 462

\bibitem[\protect\citeauthoryear{Nauenberg}{2002}]{Nauenberg2002}
      Nauenberg M., 2002, AJ, 124, 2332

\bibitem[\protect\citeauthoryear{Rabe}{1970}]{Rabe1970}
      Rabe E., 1970, in Giacaglia G.E.O., ed., Periodic Orbits, Stability and
      Resonances. D. Reidel Publ. Co., Dordrecht, p. 33

\bibitem[\protect\citeauthoryear{Rabe}{1973}]{Rabe1973}
      Rabe E., 1973, in Tapley E.D., Szebehely V., eds, Recent Advances in
      Dynamical Astronomy. D. Reidel Publ. Co., Dordrecht, p. 156

\bibitem[\protect\citeauthoryear{Robutel \& Gabern}{2006}]{RobutelandGabern2006}
      Robutel P., Gabern F., 2006, MNRAS, 372, 1463

\bibitem[\protect\citeauthoryear{S\'andor, \'Erdi \& Efthymiopoulos}{2000}]
        {Sandoretal2000}
    S\'andor Zs., \'Erdi B., Efthymiopoulos C., 2000, CeMDA, 78, 113 

\bibitem[\protect\citeauthoryear{S\'andor, Kley \& Klagyivik}{2007}]
   {Sandoretal2007}
   S\'andor Zs., Kley W., Klagyivik P., 2007, preprint
    arXiv:0706.2128v2 [astro-ph]

\bibitem[\protect\citeauthoryear{S\'andor et al.}{2004}]{Sandoretal2004}
   S\'andor Zs., \'Erdi B., Sz\'ell A., Funk B., 2004, CeMDA, 90, 127

\bibitem[\protect\citeauthoryear{Schwarz et al.}{2005}]{Schwarzetal2005}
   Schwarz R., Pilat--Lohinger E., Dvorak R., \'Erdi B., S\'andor Zs., 2005,
    Astrobiology, 5, 579

\bibitem[\protect\citeauthoryear{Schwarz et al.}{2007}]{Schwarzetal2007}
   Schwarz R., Dvorak R., Pilat--Lohinger E., S\"uli \'A., \'Erdi B., 2007,
    A\&A, 462, 1165
   
\bibitem[\protect\citeauthoryear{Tinney et al.}{2006}]{Tinneyetal2006}
   Tinney C.G., Butler R.P, Marcy G.W., et al., 2006, ApJ, 647, 594

\bibitem[\protect\citeauthoryear{Thommes}{2005}]{Thommes2005}
    Thommes E.W., 2005, ApJ, 626, 1033              

\bibitem[\protect\citeauthoryear{Tschauner}{1971}]{Tschauner1971}
    Tschauner J., 1971, CeM, 3, 189

\bibitem[\protect\citeauthoryear{Vogt et al.}{2005}]{Vogtetal2005}
    Vogt S.S., Butler P.R., Marcy W.G., et al., 2005, AJ, 632, 638
 
\end{thebibliography}
\end{document}